\documentstyle[12pt,epsf,amstex]{article}
\begin{document}
\setlength{\unitlength}{1mm}
\textwidth 15.0 true cm 
\headheight 0 cm
\headsep 0 cm 
\topmargin 0.4 true in
\oddsidemargin 0.25 true in
\input epsf

\newcommand{\beq}{\begin{equation}}
\newcommand{\eeq}{\end{equation}}
\newcommand{\be}{\begin{eqnarray}}
\newcommand{\ee}{\end{eqnarray}}
\renewcommand{\vec}[1]{{\bf #1}}
\newcommand{\vecg}[1]{\mbox{\boldmath $#1$}}
\newcommand{\grpicture}[1]
{
    \begin{center}
        \epsfxsize=200pt
        \epsfysize=0pt
        \vspace{-5mm}
        \parbox{\epsfxsize}{\epsffile{#1.eps}}
        \vspace{5mm}
    \end{center}
}

\begin{flushright}

SUBATECH--2003--06

\end{flushright}

\vspace{0.5cm}

\begin{center}

{\Large\bf  Weak Supersymmetry }

\vspace{1cm}

{\Large A.V. Smilga} \\

\vspace{0.5cm}

{\it SUBATECH, Universit\'e de
Nantes,  4 rue Alfred Kastler, BP 20722, Nantes  44307, France. }
\footnote{On leave of absence from ITEP, Moscow, Russia.}\\
\end{center}

\bigskip

\begin{abstract}
 We explore ``weak'' supersymmetric systems whose algebra involves, besides
Poincare generators, extra bosonic generators not commuting with supercharges.
This allows one to have inequal number of bosonic and fermionic 1--particle
states in the spectrum. Coleman--Mandula and Haag--Lopuszanski--Sohnius 
theorems forbid the presence of such extra bosonic charges in {\it interacting}
theory for $d \geq 3$. However, these theorems do not apply in one or two
dimensions. For $d=1$, we construct a nontrivial interacting system 
characterized by weak supersymmetric algebra. It is related to ``n--fold''
supersymmetric systems and to quasi-exactly solvable systems studied earlier. 
\end{abstract}

\section{Introduction.}

The basic defining feature of any standard supersymmetric system is double
degeneracy of all excited states. This follows from the minimal supersymmetry
algebra
  \be
\label{minsusy}
Q^2 = \bar Q^2 = 0,\ \ \ \ \ \ \{Q, \bar Q \}_+ = 2H\ ,
 \ee
where $Q$ is a complex conserved [this is a corollary of Eq.(\ref{minsusy})]
supercharge. If the superalgebra describing symmetries of the system 
includes (\ref{minsusy}) as a subalgebra, double degeneracy of all excited
levels (if supersymmetry is spontaneously broken, also the ground
 state is doubly degenerate) follows.

An interesting question is whether some other ``weak'' supersymmetric algebras
involving Poincare generators and conserved supercharges, but not including
(\ref{minsusy}) as subalgebra, are possible. At the algebraic level, the answer
is trivially positive. It is easy also to construct Lagrangians enjoying
weak supersymmetry. Indeed, the Lagrangian
  \be
\label{L4}
{\cal L} \ = \ \frac 12 \left(\partial_\mu \phi \right)^2 + i\bar \psi
\sigma_\mu \partial_\mu \psi 
 \ee
($\phi$ is a real scalar and  $\psi$ is a Weyl spinor)
is invariant with respect to supersymmetry transformations
\footnote{We use the standard Weyl  notation where   
the dotted and 
undotted indices are raised and lowered with $\epsilon^{\alpha \beta} (
\epsilon^{\dot{\alpha} \dot{\beta}} )$ and  $\epsilon_{\alpha \beta} (
\epsilon_{\dot{\alpha} \dot{\beta}} ) = - \epsilon^{\alpha \beta} (
\epsilon^{\dot{\alpha} \dot{\beta}} ) $; \ 
$\bar \psi^{\dot{\alpha}} = (\psi_\alpha)^\dagger$ and 
$\bar \psi_{\dot{\alpha}} = -(\psi^\alpha)^\dagger $; $\psi^2 = \psi_\alpha \psi^\alpha$ and
$\bar\psi^2 = \bar\psi^{\dot{\alpha}} \psi_{\dot{\alpha}}$; \ $\sigma_\mu = (1, \vecg{\sigma})$
and  $\sigma_\mu^\dagger  = (1, -\vecg{\sigma})$.}
  \be
\label{susyL4}
 \delta \phi  \ = \ \epsilon^\alpha \psi_\alpha - \bar \psi^{\dot{\alpha}}
\epsilon_{\dot{\alpha}} \ , \nonumber \\
\delta \psi_\alpha \ = \ i 
\left( \sigma ^\dagger_\mu \right)_\alpha^{\ \dot{\beta}}
\bar \epsilon_{\dot{\beta}} \partial_\mu \phi 
\ , \nonumber \\
  \delta \bar \psi^{\dot{\alpha}} \ = \ i \epsilon_{\beta }
\left( \sigma ^\dagger_\mu \right)_\beta^{\ \dot{\alpha}} \partial_\mu \phi 
\ .
 \ee
The corresponding supercharges are 
 \be
\label{QL4}
 Q_\alpha \ =\ \int d^3x \left( \sigma_\mu^\dagger \sigma_0 \psi 
\right)_\alpha \partial_\mu \phi \ , \nonumber \\
   \bar Q^{\dot{\alpha}} \ =\ \int d^3x \left( \bar \psi 
\sigma_0 \sigma_\mu^\dagger  
\right)^{\dot{\alpha}} \partial_\mu \phi \ .
 \ee
Now, $Q_\alpha$ and $\bar Q^{\dot{\alpha}}$ are conserved, but their
anticommutators involve besides $P_\alpha^{\ \dot{\alpha}}$ also extra
terms. In particular, $\{Q_\alpha, Q_\beta\} \neq 0$. The resulting 
superalgebra does not include the subalgebra (\ref{minsusy}) and the 
number of bosonic and fermionic 1--particle states might be different.
And it is: the Lagrangian (\ref{L4})  describes a free real boson (one
state $|B\rangle$ for each 3--momentum $\vec{p}$) and a free Weyl
fermion (two states $|F_\pm \rangle$). It is interesting (and important !)
that, in the sector with given  $\vec{p}$,
 the state pairing is restored for two-particle excitations and higher.
Thus, at the two--particle level, there are two boson states $|BB \rangle$
and $|F_+ F_- \rangle$ and two fermion states   $|BF_+ \rangle$ and
 $|BF_- \rangle$.
  Actually, {\it any} Lagrangian involving some number of free bosonic and some number of 
free fermionic fields is supersymmetric. There are a lot of 
such supersymmetries: each bosonic field can be mixed with each fermionic field
independently of others.
 \footnote{One of the consequencies of this is the presence of the so called
quasigoldstino branch in the spectrum of collective excitations
in quark--gluon plasma \cite{quasigold}. }
However, this is only true for free theory. As soon as the interaction 
is switched on,      supersymmetry (strong or weak) is broken. Indeed, 
nonvanishing
 $\{Q_\alpha, Q_\beta\}$ implies the presence of an extra conserved 
bosonic charge
in the representation $(1,0)$ of the Lorentz group. It is none other than the 
self-dual part of the 
fermion spin operator $S_{\alpha\beta}$. Spin is not conserved, however, in interacting theories. 
 Actually, in any theory involving
mass gap
\footnote{For massless theories, Poincare group can be extended to conformal  and super-Poincare --- to
superconformal.} and a nontrivial S--matrix, the presense of extra 
nonscalar conserving charges  is ruled out by the Coleman--Mandula theorem
\cite{Mandula}. Interacting supersymmetric theories can only involve, besides
the Poincare generators $P_\mu$, $M_{\mu\nu}$ and supercharges 
$Q^i$, central charges commuting with everything and some extra 
global symmetry generators,  which can have nontrivial commutators with $Q$ and between themselves, but they
cannot appear in this case  in  the anticommutators
of supercharges \cite{Haag}.

 This is true if the dimension of space-time is 3 or more. In two dimensions,
where scattering can be only forward or backward, the theorems 
\cite{Mandula,Haag} do not apply. In particular, one can have an infinite 
number of conserved bosonic charges (like, e.g., it is the case in the 
Sine--Gordon model). Seemingly, nothing  prevents one to have an interacting 2d theory 
enjoying a version of weak supersymmetry. 

We tried to construct one, but failed. Probably, one should try harder. What we
were able to construct is a weakly supersymmetric quantum mechanical system. 
It has two complex conserved supercharges with nontrivial anticommutators
involving besides $H$ four other bosonic generators which {\it are} not central charges
--- their commutators with supercharges and between themselves do not vanish. The boson-fermion 
degeneracy is there starting from the second excited level. But not for the
first excited level and not for vacuum. 

In Sect. 2, we describe a simplest such system --- the weak supersymmetric 
oscillator. In Sect. 3, we present a nontrivial weak supersymmetric 
Hamiltonian. We find that previously studied  quantum systems 
with so-called ``2--fold 
supersymmetry'' \cite{2fold} are in fact weak supersymmetric 
systems in disguise. We briefly discuss their relationship to quasi-exactly 
solvable models \cite{Turb}. Sect. 4 describes our failed attempt to generalize
our QM construction to $d=2$. This experience might be useful in future 
studies. Sect. 5 is reserved for discussion, conclusions, and 
acknowledgements.

\section{Weak supersymmetric oscillator}

Consider the Lagrangian 
  \be
\label{L1}
{\cal L} \ =\ \frac 12 \dot{x}^2 + i \bar \psi^\alpha \dot{\psi}_\alpha
- \frac {m^2}2 x^2 - \frac m2 (\bar \psi^\alpha \bar \psi_\alpha + \psi_\alpha
\psi^\alpha )\ ,
 \ee
$\alpha = 1,2$. It can be obtained out of the massive version of the free field theory Lagrangian (\ref{L4})
by dimensional reduction. It is invariant with respect to supersymmetry
 transformations
  \be
\label{susyL1}
\delta x &=& \bar \epsilon \psi + \bar \psi \epsilon \ , \nonumber \\
\delta \psi_\alpha &=& -i \epsilon_\alpha \dot{x} + \bar 
\epsilon_\alpha mx \ , \nonumber \\
 \delta \bar \psi^\alpha &=& i \bar \epsilon^\alpha \dot{x} - 
\epsilon^\alpha mx \ .
 \ee
The corresponding supercharges are 
  \be
\label{QL1}
Q_\alpha &=& p\psi_\alpha + imx \bar \psi_\alpha \ , \nonumber \\
\bar Q^\alpha &=& p\bar \psi^\alpha + imx \psi^\alpha \ ,
  \ee
where $p = \dot{x}$ is the bosonic canonical momentum. The canonical  Hamiltonian is 
  \be
\label{H1}
H \ =\ \frac 12 p^2 + \frac 12 m^2x^2 + \frac m2 (\bar \psi^\alpha 
\bar \psi_\alpha + \psi_\alpha \psi^\alpha )  \ .
  \ee
The quantum Hamiltonian can be written by replacing $p$ by 
$-i\partial/\partial x$ and $\bar \psi^\alpha$ by $\partial/\partial \psi_\alpha$.
Supercharges commute with the Hamiltonian. On the other hand,
  \be
\label{QZY}
\{Q_\alpha, Q_\beta \} &=& Z_{\alpha\beta} \ , \nonumber \\
\{Q_\alpha, \bar Q^\beta \} &=& \delta_\alpha^\beta 
\left(2H - Y \right)  \ ,
  \ee
where 
 \be
\label{ZY}
Z_{\alpha\beta} &=&  m(\psi_\alpha \bar \psi_\beta  + \psi_\beta \bar 
\psi_\alpha  )\ , \nonumber \\
 Y &=& \frac m2  (\bar \psi^\alpha  \bar \psi_\alpha + \psi_\alpha \psi^\alpha )\ .
  \ee
The operators $Z_{\alpha\beta}$ and $Y$ commute with $H$ and with each other, 
but  the commutators
 \be
\label{comQZY}
\left[ Q_\alpha , Z_{\beta\gamma} \right]  &=&  m \left( \epsilon_{\alpha\beta} Q_\gamma +
\epsilon_{\alpha\gamma} Q_\beta \right) \ , \nonumber \\
   \left[ {\bar Q}_\alpha , Z_{\beta\gamma}  \right]  &=& m \left( \epsilon_{\alpha\beta} 
\bar Q_\gamma +
\epsilon_{\alpha\gamma} \bar Q_\beta \right) \ , \nonumber \\
\left[ Q_\alpha, Y \right] &=& m \bar Q_\alpha\ , \nonumber \\
 \left[ \bar Q_\alpha, Y \right] &=&  m  Q_\alpha
  \ee
and also 
  \be
 \label{comZZ}
\left[ Z_{\alpha\beta}, Z_{\gamma\delta} \right] \ =\ 
m \left( \epsilon_{\beta\gamma} Z_{\alpha\delta} +   \epsilon_{\alpha\gamma} Z_{\beta\delta}
+ \epsilon_{\beta\delta} Z_{\alpha\gamma} + \epsilon_{\alpha\delta} Z_{\beta\gamma} \right)
  \ee   
are nontrivial. The algebra can be presented in a little bit more convenient
form if introducing $S_\alpha = Q_\alpha - \bar Q_\alpha,\ \ 
\bar S^\alpha = Q^\alpha + \bar Q^\alpha$. Then
   \be
\label{comSZY}
\{S_\alpha, \bar S^\beta \} &=& 4H \delta_\alpha^\beta - 2Y \delta_\alpha^\beta
+ 2 Z_\alpha^\beta \ , \nonumber \\
\left[ S_\alpha , Z_{\beta\gamma} \right] &=& m \left( \epsilon_{\alpha\beta} S_\gamma +
\epsilon_{\alpha\gamma} S_\beta \right)\ , \nonumber \\
 \left[\bar S^\alpha , Z_{\beta\gamma} \right] &=& m \left(\delta^\alpha_\beta  
\bar S_\gamma +
\delta^\alpha_\gamma  \bar S_\beta \right)\ , \nonumber \\
\left[ S_\alpha, Y \right] &=& - m S_\alpha,\  \nonumber  \\
\left[ \bar S^\alpha, Y \right] &=&  m \bar S^\alpha\ ,
  \ee
to which the commutator (\ref{comZZ}) should be added. The subalgebra (\ref{comZZ}) is none other
than $sl(2)$ , which can be readily seen by identification 
$$ Z_{11} \equiv 2im\sigma_+,\ \  Z_{22} \equiv 2im\sigma_-,\ \ Z_{12} = Z_{21} \equiv -m\sigma_3 \ .$$ 
All other commutators and the anticommutator $\{S_\alpha, S_\beta \}$ vanish. 
 
It is not difficult to find the spectrum of $H$. The eigenstates are
 \be
\label{states}
\Phi_\pm^n &=&  \frac 1 {\sqrt{2}}|n\rangle \left( 1 \pm \frac 12 \psi_\alpha \psi^\alpha 
\right);   \ \ \ \ \ \ \Phi_\alpha^n = \psi_\alpha |n\rangle \ ,
 \ee
where $|n\rangle $ are the bosonic oscillator eigenstates. Their energies are
  \be
\label{Ener}
E_-^n \ = \ \left( - \frac 12 + n \right)m\ , \ \ 
E_\alpha^n \ = \ \left(  \frac 12 + n \right)m\ , \ \  
E_+^n \ = \ \left(  \frac 32 + n \right)m\ .
  \ee
The spectrum is drawn in Fig. 1. We see that there is one vacuum state
(its energy can be brought to zero by adding the constant $m/2$ to the 
Hamiltonian, but for the weak supersymmetric systems with algebra 
(\ref{comSZY}), $E_{\rm vac} = 0$ is an {\it option} rather than requirement).
There are three first excited states: a bosonic and two fermionic. Starting from
the second excited state, there are 2 bosonic and 2 fermionic states at each level.

\begin{figure}
   \begin{center}
        \epsfxsize=350pt
        \epsfysize=150pt
        \vspace{25mm}
        \parbox{\epsfxsize}{\epsffile{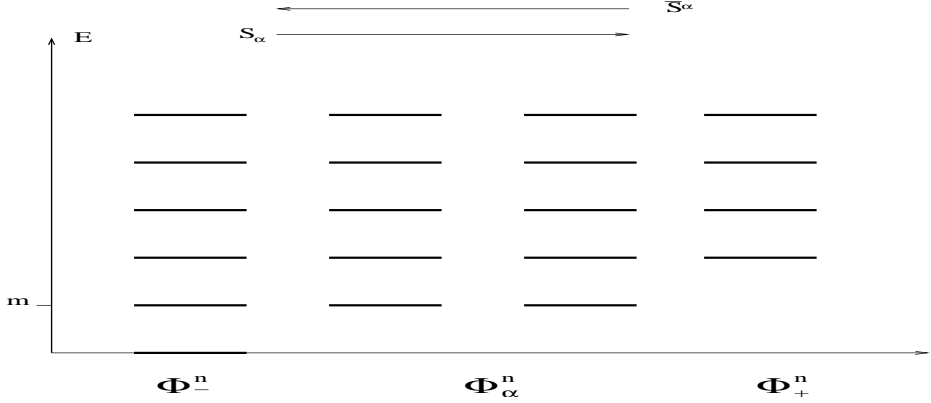}}
        \vspace{-5mm}
    \end{center}
\caption{Spectrum of weak supersymmetric oscillator.}
\end{figure}

The eigenvalues of the operator $Y$ is $-m$ for the leftmost tower, 0 for two central and
$+m$ for the rightmost one. In other words, the operator $Y/m$ (or rather
 $Y/m + 1$)
 plays the role of the fermionic
charge. The operators $Z_{\alpha\beta}$ annihilate the states $\Phi_\pm^n$ 
while the states $\Phi_\alpha^n$ form doublet representations of the $sl(2)$ algebra (\ref{comZZ}). 

 To acquire further insights, it is instructive to write the action of the operators $S_\alpha,
\bar S^\alpha$ on the states (\ref{states}):
  \be
\label{Snasost}
S_\alpha \Phi_-^n = 2\sqrt{mn} \Phi^{n-1}_\alpha,\ \ \ S_\alpha \Phi^n_\beta = 
-2\sqrt{mn} \epsilon_{\alpha\beta} \Phi_+^{n-1},\ \ S_\alpha \Phi^n_+ = 0\ , \nonumber \\
\bar S^\alpha \Phi^n_- = 0, \ \ \ 
 \bar S^\alpha \Phi^n_\beta = 
2\sqrt{m(n+1)} \delta_{\beta}^\alpha \Phi_-^{n+1}, \ \ \ \bar S^\alpha \Phi_+^n = 2\sqrt{m(n+1)}
\epsilon^{\alpha\beta} \Phi^{n+1}_\beta\ .
 \ee
We see that $S_\alpha$ annihilates the states from the rightmost column and brings the states from
the left columns to the right. The action of $\bar S^\alpha$ is opposite.

Now, one can divide all eigenstates in two sets: {\it (i)} the states $\Phi^n_-$ and $\Phi_1^n$
and {\it (ii)} the states $\Phi_2^n$ and $\Phi_+^n$. The states from the subset {\it (i)} form the Hilbert
space of the ${\cal N} = 1$ supersymmetric oscillator, with $S_1$ playing the role of the supercharge.
The same applies to the subset {\it (ii)} with the supercharge $S_2$. 
For  ordinary  ${\cal N} = 2$ supersymmetric quantum mechanics, the Hilbert space can also be divided
into two  ${\cal N} = 1$ subspaces, but the specifics of a weak supersymmetric system is that two
sets of states are shifted with respect to each other, i.e. the Hamiltonian for the right subset
differs from the Hamiltonian for the left subset by a constant.

    \section{A class of interactive weak supersymmetric systems.}
Consider the Lagrangian
 \be
\label{L1I}
{\cal L} \ =\ \frac {\dot{x}^2}2 + i \bar \psi^\alpha \dot{\psi}_\alpha - \frac {V^2}2 - \frac {V'}2
\left(\psi^2 + \bar \psi^2 \right) - \frac {B'}2 \psi^2 \bar \psi^2\ , 
  \ee
$$ B \ =\ \frac {V'-C}{2V}\ ,$$
where $V(x)$ is an arbitrary function and $C$ is an arbitrary constant. One can observe that the corresponding
action is invariant with respect to the supersymmetry transformations
  \be
\label{trans}
\delta x &=& \bar \epsilon \psi + \bar \psi \epsilon \ , \nonumber \\
\delta \psi_\alpha &=& - i\epsilon_\alpha \dot{x} + 
\bar \epsilon_\alpha (V + B \psi^2) - 2B\left[ (\bar\psi \psi) \epsilon_\alpha
+ (\bar \psi \epsilon) \psi_\alpha \right] \ , \nonumber \\
\delta \bar\psi^\alpha &=& i\bar \epsilon^\alpha \dot{x} - \epsilon^\alpha (V + B \bar\psi^2) 
 - 2B\left[ \bar \epsilon^\alpha (\bar\psi \psi)
+  \bar\psi^\alpha (\bar \epsilon \psi ) \right] \ .
  \ee
When $V(x) = mx$ and $C=m$, Eq.(\ref{L1I}) is reduced to the oscillator Lagrangian (\ref{L1}) considered above.
The first four terms in Eq.(\ref{L1I}) represent a rather natural generalization of Eq.(\ref{L1}), like
in Witten's supersymmetric quantum mechanics \cite{Wit}. In our case, we are obliged to add also a 4-fermion term in 
the Lagrangian and 
extra nonlinear terms in  the transformation law. 

The canonical classical supercharges and Hamiltonian are
  \be
\label{Qclas}
Q_\alpha &=& p\psi_\alpha + iV \bar\psi_\alpha + iB \psi^2 \bar\psi_\alpha \ , \nonumber \\
\bar Q^\alpha &=&  p\bar \psi^\alpha + iV \psi^\alpha + iB \bar\psi^2 \psi^\alpha \ ,
  \ee
  \be
\label{Hclas}
H^{\rm cl} \ =\ \frac {p^2}2 + \frac {V^2}2 + \frac {V'}2 \left( \psi^2 + \bar\psi^2 \right) + \frac {B'}2 \psi^2
\bar\psi^2\ .
  \ee
The Poisson brackets $\{Q_\alpha, H \}_{\rm P.B.}$ and $\{\bar Q^\alpha, H \}_{\rm P.B.}$ vanish. 
A certain care is required when quantizing this theory. We want to fix the ordering ambiguities in $Q$ and $H$
 such  that classical supersymmetry were  not spoiled at the quantum level. An experience acquired by fiddling
with supersymmetric $\sigma$--models and gauge theories \cite{quant} teaches us that proper quantum {\it supercharges}
should be obtained by Weyl ordering procedure from the classical expressions (\ref{Qclas}). In other words, (\ref{Qclas})
should be Weyl symbols of the quantum supercharges. This does not necessarily apply to the Hamiltonian.
Indeed, if choosing the quantum Hamiltonian such that (\ref{Hclas}) represents the Weyl symbol of $\hat{H}^{\rm qu}$,
the Weyl symbol of the commutator $[\hat{Q}_\alpha, \hat{H} ]$ would be given by the Moyal bracket \cite{Moyal,quant}
of $Q_\alpha^{\rm cl}$ and $H^{\rm cl}$, 
  \be
\label{Moyal}
\left[ \hat{Q}_\alpha, \hat{H} \right]_W \ =\ \{Q_\alpha, H \}_{\rm M.B.} \ = \nonumber \\
2 {\rm sh} \left\{ \frac 12 \left[ \frac {\partial^2}{\partial \Psi_\alpha \partial \bar \psi^\alpha}
- \frac {\partial^2}{\partial \psi_\alpha \partial \bar \Psi^\alpha} + i\left( 
\frac {\partial^2}{\partial q \partial P} - \frac {\partial^2}{\partial Q \partial p} \right) \right] \right\}
\nonumber \\
\left. Q_\alpha^{\rm cl} (\bar\psi, \psi; p,q) H^{\rm cl} (\bar\Psi, \Psi; P,Q) \right|_{ \bar\Psi = \bar\psi, \Psi = \psi; 
P=p, Q=q}
= \ iBB' \psi_\alpha \neq 0\ .
  \ee
To compensate that, we should add to $H^{\rm cl}$ the term $B^2/2$. The quantum supercharges and Hamiltonian thus 
obtained are
    \be
\label{Qquant}
\hat{Q}_\alpha &=& {p}\psi_\alpha + iV {\bar\psi}_\alpha + iB \left( \psi^2 {\bar\psi}_\alpha 
+ \psi_\alpha \right)\ , \nonumber \\
\hat{\bar Q}^\alpha &=&  {p}{\bar \psi}^\alpha + iV \psi^\alpha + iB \left(  \psi^\alpha {\bar\psi}^2
- {\bar\psi}^\alpha \right)  
  \ee
and
  \be
\label{Hquant}
\hat{H} \ =\ \frac {{p}^2}2 + \frac {V^2}2 + \frac {V'}2 \left( \psi^2 + {\bar\psi}^2 \right) + \frac {B'}2 
\left( \psi^2
{\bar\psi}^2 - 2\psi {\bar\psi} + 1 \right) + \frac {B^2}2 + \frac C2\ ,
  \ee
where we added for convenience the constant $C/2$ in the Hamiltonian. Direct calculation of the commutators 
(or, which is simpler, of the Moyal brackets of the classical expressions) leads to a remarlable conclusion:
the algebra (\ref{comSZY}, \ref{comZZ}) derived for the oscillator is valid also in the interactive case, with $m \to C$, 
$H \to H -C/2$ and 
$Z_{\alpha\beta}, Y$ having the same form (\ref{ZY}) as before.

As earlier, the quantum states can be divided into three classes: {\it (i)} the states $|-\rangle \propto
1 - \psi^2/2$, {\it (ii)} the states $|\alpha \rangle \propto \psi_\alpha$ (they are present in two copies as 
the Hamiltonian (\ref{Hquant}) does not feel the index $\alpha$)
 and {\it (iii)} the states 
$|+ \rangle \propto 1 + \psi^2/2$. These  states are characterized by a definite value 
of the ``fermion charge'' $Y$ : $Y_\pm = \pm m$ and $Y_\alpha
= 0$. In each such sector, we have an ordinary Schr\"odinger equation with the potentials
  \be
\label{Upm}
U_- &=& \frac 12 (W_-^2 - W_-' ) \ ,\nonumber \\
U_\alpha &=& \frac 12 (W_-^2 + W_-' ) \ =\  \frac 12 (W_+^2 - W_+' ) + C\ , \nonumber \\
U_+ &=& \frac 12 (W_+^2 + W_+' ) + C \ ,
  \ee
where 
 \be
\label{Wpm}
W_\pm = V \pm B \ .
  \ee
It is clear now that we are dealing with two superimposed ordinary Witten's SQM systems.  The states 
$|- \rangle$ and $|1 \rangle$ are described by such system with superpotential $W_-$ and the states 
$|2 \rangle$ and $|+ \rangle$ - by the  system with superpotential $W_+$, with the constant $C$ added to the Hamiltonian. 
 Excited
states are mostly 4-fold degenerate as for usual ${\cal N} = 2$ SQM.

The ground state is not necessarily degenerate. If $\exp\{ - \int W_-(x) dx \}$ is normalizable, this
(being multiplied by $1 - \psi^2/2$)  determines the wave function of the unique vacuum state. With the chosen
normalization of the Hamiltonian  [the term $C/2$ in Eq.(\ref{Hquant}) !] it has zero energy.
Further, if  $\exp\{ - \int W_+(x) dx \}$ is normalizable, there is also a unique zero-energy ground state for  Witten's
Hamiltonian with superpotential $W_+$. Thus, we obtain a state in the sector 
$|2 \rangle$ with energy $C$. Due to $|2 \rangle  \leftrightarrow |1 \rangle$ and $|1 \rangle \leftrightarrow |- \rangle $
 degeneracies,  we have altogether {\it three} states with energy $C$ at the first excited level, and the picture
is the same as for the oscillator (see Fig.1).

We have obtained free of charge a wide class of {\it quasi-exactly solvable} \cite{Turb} 
 potentials $U_-(x)$ for which the energy
of the ground state ($E_0 =0$) and of the first excited state $E_1 = C$ are exactly known. They depend on an arbitrary
function $V(x)$ and an arbitrary constant $C$  with a certain restriction:
 both  $\exp\{ - \int W_-(x) dx \}$ and $\exp\{ - \int W_+(x) dx \}$ should be normalizable. Probably, the simplest
nontrivial choice is $V(x) = mx + \alpha x^3$ with $C = m$ ($m, \alpha  > 0$). 
Note that the potential $U_-(x)$   is not polynomial in this case.

The potentials $U_\pm$ in Eqs.(\ref{Upm}, \ref{Wpm}) were discussed before (see Eq.(4.18) in Ref.\cite{japoncy}
and Eq.(50) in Ref.\cite{AndSok} ) in association with the so called 2--fold supersymmetry construction
developped in \cite{2fold}. $N$--fold supersymmetry is a supersymmetry where supercharges are not linear
in momentum, but present polynomials of power $N$. In our case, one can define the quadratic in $p$
operator
   \be
\label{Qcal}
{\cal Q} \ = \ S_1 S_2
  \ee
with the action ${\cal Q}|+\rangle = {\cal Q} |\alpha \rangle = 0$ and ${\cal Q} |-\rangle = |+ \rangle$.
$\bar {\cal Q}$ acts in the opposite direction:  
$\bar {\cal Q} |+ \rangle = |- \rangle,\ \   \bar {\cal Q}| \alpha \rangle = \bar {\cal Q} |- \rangle = 0$.
The operators ${\cal Q}, \bar {\cal Q}$ commute with $H$ (as $S_{1,2}$ and $\bar S^{1,2}$ do). If disregarding
the states $|\alpha \rangle$ annihilated by both  ${\cal Q}$ and $ \bar {\cal Q}$ and considering
 only the sectors $|+ \rangle$ and $|- \rangle$, one can deduce
  \be
\label{QQH}
 \{{\cal Q}, \bar{\cal Q} \} \ =\ 16H(H-C) \ .
   \ee
The quadratic polynomial of $H$ appearing on the right side is characteristic of 2--fold supersymmetry.
The full algebra of the weak supersymmetry (\ref{comZZ}, \ref{comSZY}) displays itself only if the 
``central'' sector $|\alpha \rangle$ is brought into consideration.

  \section{A generalization that failed.}

As was discussed in the Introduction, it would be very interesting to construct a nontrivial 
$2D$ weak supersymmetric model. In this section, we explain why a naive generalization of the model 
(\ref{L1I}) to the $2D$ case fails. Let us consider only the fermion part in (\ref{L1I}) and
assume $x$ to be constant. Such a restricted action is still invariant with respect to
$\delta \psi_\alpha = \ldots, \delta \bar\psi^\alpha = \ldots$ as dictated by Eq.(\ref{trans})
Consider only the part in $\delta {\cal L}$ involving the time derivatives $\dot{\psi}_\alpha$ and 
$\dot{\bar\psi}^\alpha$ and cubic in $\psi$. It is determined by the variation of the kinetic term 
$i\bar\psi^\alpha \dot{\psi}_\alpha$ coming from the bits in $\delta {\psi_\alpha}$ and 
$\delta{\bar\psi^\alpha}$ that are quadratic in $\psi, \bar\psi$. Such a variation represents a total
time derivative as it should. 

 Let us now try to generalize this to $d=2$. Consider the fermion kinetic term
\footnote{The convention in this section is $\bar\psi = \psi^\dagger \gamma^0, 
\gamma_\mu \gamma_\nu = g_{\mu\nu} - \epsilon_{\mu\nu} \gamma^5$ (e.g. $\gamma_0 = \sigma^1,
\gamma_1 = i\sigma^2, \gamma_5 = \sigma^3$).}
  \be
\label{Lfkin}
{\cal L}_f \ =\ i\bar\psi \gamma_\mu \partial_\mu \psi
  \ee
and let us require it to be invariant up to a total derivative with respect to the transformation
  \be
\label{var2D}
\delta \psi_\alpha &=& (\bar \psi \epsilon) \psi_\alpha + A  (\bar \psi \gamma_5 \epsilon)( 
\gamma_5 \psi)_\alpha + B  (\bar \psi \gamma_\nu \epsilon)( 
\gamma_\nu \psi)_\alpha \ , \nonumber \\
\delta \bar\psi^\alpha &=& D\bar\psi^2 \epsilon^\alpha
  \ee
with arbitrary constants $A,B,D$ (only the terms $\propto \epsilon$ are displayed, the terms
$\propto \bar\epsilon$ to be restored by hermiticity). Consider the terms $\propto \partial_\mu \bar\psi$
in $\delta {\cal L}_f$:
   \be
\label{vardpsi}
\delta {\cal L}_f \ \propto \ (\bar\psi \gamma_\mu \psi)(\partial_\mu \bar\psi \epsilon) +
A (\bar\psi \gamma_\mu \gamma_5 \psi)(\partial_\mu \bar\psi  \gamma_5 \epsilon)
+ B (\bar\psi  \psi)(\partial_\mu \bar\psi \gamma_\mu \epsilon) \nonumber \\
-B \epsilon_{\mu\nu} (\bar\psi \gamma_5\psi)(\partial_\mu \bar\psi \gamma_\nu \epsilon)\ .
   \ee
Introduce the charge congugation matrix $C= i \sigma^2$ satisfying the properties
  \be
\label{C}
C^T = -C,\ \ C^2 = -1,\ \ C\gamma_\mu = -\gamma_\mu^T C
 \ee
and rewrite Eq.(\ref{vardpsi}) in terms of the structures $\bar\psi C \partial_\mu \bar\psi, \ 
\bar\psi \gamma_\nu C \partial_\mu \bar\psi $ and $\bar\psi \gamma_5 C \partial_\mu \bar\psi $ 
using the Fierz identity 
 \be
\label{Firc}
\bar\psi_\alpha \partial_\mu \bar\psi_\gamma \ =\  -\frac 12 \left[
C_{\alpha\gamma} (\bar\psi C \partial_\mu \bar\psi) + 
(C \gamma_\nu)_{\alpha\gamma} (\bar\psi \gamma_\nu C \partial_\mu \bar\psi) + 
(C \gamma_5)_{\alpha\gamma} (\bar\psi \gamma_5 C \partial_\mu \bar\psi)\right] \ .
  \ee
If we wish the variation (\ref{vardpsi}) to be a total derivative, 
only the structure $\bar\psi C \partial_\mu \bar\psi
= \frac 12 \partial_\mu [\bar\psi C  \bar\psi ] $ in $\delta {\cal L}_f$ 
is allowed while the coefficients of the structures   $\bar\psi \gamma_\nu C \partial_\mu \bar\psi $
and  $\bar\psi \gamma_5 C \partial_\mu \bar\psi $ should vanish --- the matrices $\gamma_\nu C $
and $\gamma_5 C$ are symmetric and   $\bar\psi \gamma_\nu C  \bar\psi $
and  $\bar\psi \gamma_5 C  \bar\psi $ vanish identically. 

The statement is that it {\it is} not possible to suppress unwanted structures with any choice 
of the constants $A,B$. 

  \section{Discussion.}
Our original  motivation was the quest for nontrivial supersymmetric systems with mismatch
between bosonic and fermionic degrees of freedom. Let us note here that, while it is difficult
 to find such {\it field theory} systems, their presence in quantum mechanics 
 was known for a long time. Most popular SQM systems (Witten's quantum mechanics, standard
supersymmetric $\sigma$ models, etc) have an equal number of bosonic and fermionic phase space 
coordinates. But  the SQM system describing planar motion in transverse 
magnetic field involves two pairs of bosonic variables and only one pair of fermionic variables.
A class of nonstandard ``symplectic''  ${\cal N} = 2$ supersymmetric $\sigma$ models involving $3r$ bosonic 
variables and $2r$ fermionic variables ($r$ is an integer) was constructed in Ref.\cite{jaIv}. The Diaconescu-Entin
${\cal N} =4$ symplectic $\sigma$ model \cite{DE} generalized in \cite{N2} involves $5r$ bosonic
and $4r$ fermionic variables. An industrial method to construct SQM models where the number of bosonic
variables is {\it less} than the number of fermionic ones was suggested in \cite{Losev}. In SQM, an
equal number  bosonic and fermionic degrees of freedom {\it is} not required by supersymmetry. What is
required is the equal number of bosonic and fermionic quantum states. But in field theory, any bosonic
or fermionic dynamical field correspond to an asymptotic state (a particle), and bosons and fermions should
normally be matched.

We notice that this matching can be absent if relaxing the requirement that the anticommutator
of supercharges involves only the Hamiltonian, momentum, and central charges. A lot of free
weak supersymmetric models can be written, but, for $d > 3$, {\it interactive} weak supersymmetric theories
do not exist. This follows from the Haag--Lopuszanski--Sohnius theorem. This theorem does not apply to 
2 dimensions, however, and the existence of interactive      weak supersymmetric theories cannot be ruled
out. The fact that our quest was not successful leaves two options:
 \begin{itemize} 
\item Maybe one should just try harder and such systems will eventually be found. Pehaps, it is reasonable
to look at supersymmetric generalizations of exactly solvable $2D$  models with an infinite number of conservation
laws (such bosonic systems do not exist for $d > 3$ due to Coleman-Mandula theorem,
which is relative to the HLS theorem). 
However, in known such supersymmetric generalizations \cite{Sine-Gord}, there is no mismatch.   
 \item Maybe such models do not exist, indeed, in which case one should be able to prove 
a strong version of the  Haag--Lopuszanski--Sohnius theorem.
  \end{itemize}
Both possibilities look very interesting and only future studies will show which of them is
realized.

The main {\it positive} result of this paper is the system (\ref{L1I}) which enjoys 
a weak supersymmetry algebra (\ref{comZZ}), (\ref{comSZY}). It describes quantum systems
which were studied before, but from a different perspective. It would be interesting to construct
and study other, more complicated weak supersymmetric models, especially the models involving several
bosonic degrees of freedom. This would allow one to construct new examples of multidimensional
quasi-exactly solvable models.

\section*{Acknowledgements}
I am indebted to N. Dorey, M. Henneaux,  N. Nekrasov, M. Plyushchay, M. Shifman, and A. Vainshtein
for illuminating discussions and correspondence.

\end{document}